\documentstyle{elsart}

\newcommand \fr[2] {\mbox{${#1\over #2}$}}
\newcommand \FR[2] {\mbox{$\,\displaystyle{\strut\displaystyle #1 \over
 \strut\displaystyle #2}\,$}}

\begin{document}
\begin{frontmatter}

\title{ON-SHELL2: FORM based package for the calculation of two-loop
           self-energy single scale Feynman diagrams occurring
           in the Standard Model}


\author[A]{J.Fleischer },
\author[B,C]{M.~Yu.~Kalmykov},

\thanks[A]{E-mail: fl@physik.uni-bielefeld.de}
\thanks[B]{E-mail: misha@physik.uni-bielefeld.de}
\thanks[C]{On leave of absence from JINR, 1141980 Dubna (Moscow Region) Russia}

\address{Fakult\"at f\"ur Physik, Universit\"at Bielefeld,
D-33615 Bielefeld, Germany}

\begin{abstract}
A FORM based package ({\bf ON-SHELL2}) for the calculation of two loop
self-energy diagrams with one nonzero mass in internal lines and the external
momentum on the same mass shell is elaborated. The algorithm, based on
recurrence relations obtained from the integration-by-parts method, allows us
to reduce diagrams with arbitrary indices (powers of scalar propagators)
to a set of master integrals. The SHELL2 package is used for the 
calculation of special types of diagrams. Analytical results for 
master integrals are collected. 
\end{abstract}



\end{frontmatter}
\input epsf.tex

\section{Introduction}
\vspace*{-0.5pt}

High experimental accuracy achieved in the last years
allows to test the Standard Model on the level of quantum corrections. 
Therefore to improve the accuracy of predictions within this model
is of urgent need. The calculation of 
mass-dependent radiative corrections is complicated but can be performed 
to a large extent by using computer algebra.
In recent years many algorithms have been developed 
and huge program packages were elaborated for this purpose
(for a review of existing packages see Ref.\cite{review}). 
Due to different mass scales of the 
particles in the Standard Model the method of asymptotic expansion
\cite{asymptotic} - the expansion of Feynman diagrams w.r.t.
the ratio of different scale parameters - is becoming more and more
popular. The calculation of radiative corrections to low energy processes e.g.,
in particular those with light external fermions, in many cases reduces
to the calculation of self-energy diagrams with external momentum at 
different scales. This is one of the reasons why the evaluation of
two-loop self-energy diagrams is worth special attention. From the point
of view of approximation methods  two-loop self-energy diagrams can be 
divided into several classes:

\begin{itemize}
\item
Only one non-zero mass enters internal lines and the 
external momentum is on the same mass shell. 
The calculation of diagrams of this type occurring in QED and QCD has
been implemented
\footnote{One of the first calculations of this type
for the Standard Model and QED was performed in Refs.\cite{S-M,1david}.}
as the package SHELL2 \cite{SHELL2}.
\item 
There are heavy particles in internal lines and the external
momentum is on the mass-shell of a light particle \cite{small1,small2,small3}.
Diagrams of this type occurring in the Standard Model have been
collected in the package TLAMM \cite{TLAMM}.
\item
All internal particles are light or massless and the external momentum
is on a heavy mass shell (see, for example, Ref. \cite{large}).
\item
Several different heavy masses occur in internal lines and the external
momentum is on the mass shell of a heavy particle \cite{CS,AK}. 
\item
Diagrams close to thresholds 
(see Ref.\cite{threshold} and references therein).
\end{itemize}

A general recipe of reduction of arbitrary two-loop self-energy
diagrams to a set of master integrals has been suggested by Tarasov  
\cite{Tarasov1}. The algorithm was implemented in FORM and then 
in MATHEMATICA \cite{OS} . However, this method suffers from 
the drawback that in processing the reduction of scalar master 
integrals with shifted dimension to the generic dimension of space-time, 
powers of $1\over\varepsilon$ may arise which require the 
expansion of master integrals as series in $\varepsilon$, 
which is a difficult task. This problem is avoided in our approach.

We present a FORM \cite{FORM} based package that allows
to calculate arbitrary two-loop self energy diagrams 
with one non-zero mass and the external
momentum on the (nonvanishing) mass shell. Our algorithm concerning the V-type
diagrams is very similar to the one described in Ref.\cite{similar}. 

The paper is organized as follows. In Sect.2 the full set
of recurrence relations is presented which allows to express the initial diagrams 
with scalar product in the numerator in terms of diagrams with 
positive ($>1$) indices. Sect.3 is devoted to the description of how to use
the package. In appendix A (appendix B) we give all needed recurrence 
relations to reduce the scalar F-prototypes (V-prototypes) with
arbitrary positive indices to a set of master-integrals.
In appendix C the analytical results for all integrals shown in 
Fig.\ref{joint}  with indices 1 are collected.  
Even though not all of these are considered as master integrals, they
can be used for comparison. We are working in Euclidean 
space-time with dimension $N = 4 - 2  \varepsilon$.

\section {The recurrence relations.}

\begin{figure}[ht]
\centerline{\vbox{\epsfysize=155mm \epsfbox{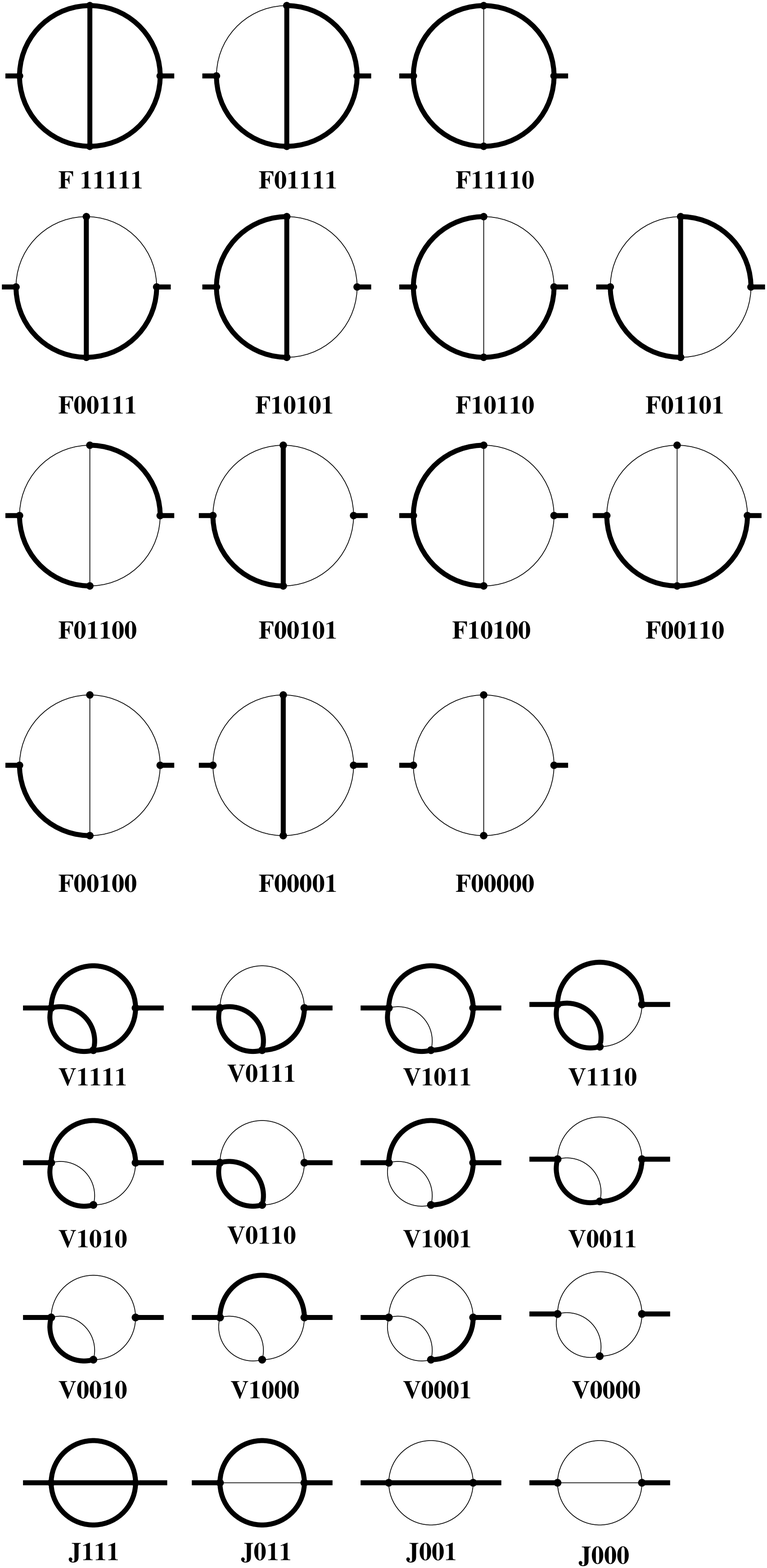}}}
\caption{\label{joint} The F, V  and J topologies.
Bold and thin lines correspond to the mass and
massless propagators, respectively.}
\end{figure}

\begin{figure}[ht]
\centerline{\vbox{\epsfysize=30mm \epsfbox{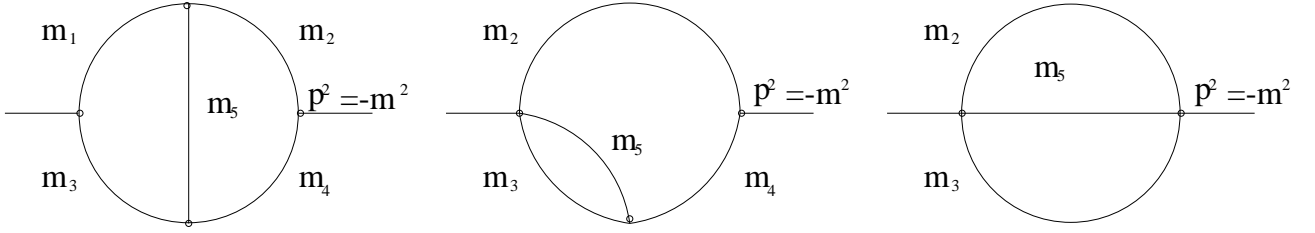}}}
\caption{\label{notation} Notations used in present paper.}
\end{figure}

The full set of two-loop self energy diagrams with one mass and external
momentum on the same mass shell is given in Fig.\ref{joint}. We distinguish
three  basic topologies which in accordance with  notations in
Ref.\cite{Tarasov1} we call F, V and J prototypes with five, four and
three lines, respectively. Our notation is given in Fig.\ref{notation}. 
The diagrams implemented in the package SHELL2  (F01101, F00110, V1110,
V0011, V1000 in our notation) 
and those considered in detail in Refs.\cite{alvladim}-\cite{program} 
(F00000, V0000, J001, J000) are not discussed here 
\footnote{The procedures for the calculation of all diagrams of the topologies
shown in Fig.\ref{joint} are implemented in our package.}.

\begin{figure}[ht]
\centerline{\vbox{\epsfysize=60mm \epsfbox{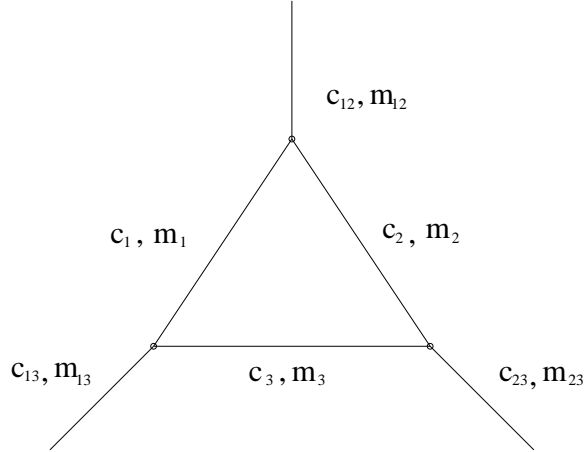}}}
\caption{\label{triangle} ``Triangle'' rule. }
\end{figure}

The general prototype involves arbitrary integer powers of the scalar 
denominators $c_L = k_L^2 + m_L^2$ 
\footnote{Their explicit expressions are
$c_1 = k_1^2 + m_1^2,~~~c_2 =  k_2^2 + m_2^2, 
~~~c_3 = (k_1-p)^2+m_3^2,~~~c_4 = (k_2-p)^2+m_4^2,
~~~c_5 = (k_1-k_2)^2 + m_5^2$}.  
Their powers 
$j_L$ are called indices of the lines. The mass-shell condition 
for the external momentum now is $p^2=-m^2$. 
Any scalar products of the momenta in the numerator arising from
projection or expansion are reduced to powers of the scalar propagators
(in case of V and J topologies the corresponding 
lines are added). Thus, the indices may sometimes become negative. 
Recurrence relations are derived via the integration-by-parts 
method \cite{rec} and applied to the massive case as in
Ref.\cite{kotikov}. They allow to reduce all lines with negative indices 
to zero and the positive indices to one or zero.  Further we use the 
shorthand notation $\{123\}$ of Ref.~\cite{avdeev} to denote 
the relation for the triangle formed of lines $1$, $2$, and $3$:
\newcommand\eqnum[1] {\eqno{#1}}

$$ \int \frac{d^N k }{c_1^{j_1} c_2^{j_2} c_3^{j_3}}
\Big( N -2 j_1 -j_2 -j_3 +j_1 \frac{2 m_1^2}{c_1}
+j_2\frac{m_1^2+m_2^2-m_{12}^2+c_{12}-c_1}{c_2}
$$
$$
+j_3\frac{m_1^2+m_3^2-m_{13}^2+c_{13}-c_1}{c_3} \Big) = 0 ,
\eqnum{\{123\}} $$

\noindent where a double index like $\{ 12 \}$ refers to a line that starts at
the point where lines $1$ and $2$ meet (see Fig.\ref{triangle}). 
For an external line on the mass shell, the value of $c_L$ is equal 
to zero. 

\subsection{F-topology}

To exclude the numerator (for example, $j_1 <0,~~~c_1^{|j_1|}$ in the
numerator) of F-type diagrams  we use the following set of 
recurrence relations:

\begin{enumerate}
\item $j_5 \neq 1$

\begin{eqnarray}
\{245\} &&  
\frac{j_5}{c_5} c_1 = 
- \frac{j_2}{c_2} 2 m_2^2
- \frac{j_4}{c_4} \left( m_4^2 +m_2^2 -m^2 - c_2 \right)
\nonumber \\
&& 
- \frac{j_5}{c_5} \left( m_5^2 +m_2^2 -m_1^2 - c_2 \right)
- N+ 2 j_2+j_4+j_5 , 
\nonumber 
\end{eqnarray}

\item $j_2 \neq 1$
\begin{eqnarray}
\{524\} &&  
\frac{j_2}{c_2} c_1 = 
- \frac{j_5}{c_5} 2 m_5^2
- \frac{j_4}{c_4} \left( m_5^2 +m_4^2 -m_3^2 + c_3 - c_5 \right)
\nonumber \\
&& 
- \frac{j_2}{c_2} \left( m_5^2 +m_2^2 -m_1^2 - c_5 \right)
- N+ 2 j_5+j_4+j_2 ,
\nonumber 
\end{eqnarray}

\item $j_3 \neq 1$
\begin{eqnarray}
\{245\} + \{135\} &&  
\frac{j_3}{c_3} c_1 = 
\frac{j_3}{c_3} \left( m_3^2 +m_1^2 -m^2 \right)
\nonumber \\
&& 
+ \frac{j_4}{c_4} \left( m_2^2 +m_4^2 -m^2 \right)
\nonumber \\
&& 
- \frac{j_4}{c_4} c_2
+ \frac{j_1}{c_1} 2 m_1^2
+ \frac{j_2}{c_2} 2 m_2^2
+ \frac{j_5}{c_5} 2 m_5^2
\nonumber \\
&& 
+ 2N -2j_1 -2j_2 -j_3 -j_4 -2j_5 , 
\nonumber 
\end{eqnarray}

\item $j_2 = j_3 = j_5 = 1$
\begin{eqnarray}
\{315\} &&
N -2j_3 -j_1 -j_5 = 
- \frac{j_1}{c_1} \left( m_1^2 +m_3^2 -m^2 - c_3 \right)
\nonumber \\
&& 
- \frac{j_5}{c_5} \left( m_3^2 +m_5^2 -m_4^2 + c_4 - c_3 \right)
- \frac{j_3}{c_3} 2 m_3^2 ,
\nonumber 
\end{eqnarray}
\end{enumerate}

\noindent
where both sides of these relations are understood to be multiplied by
\newline
$
\int \frac{d^N k_1 d^N k_2}{c_1^{j_1} c_2^{j_2} c_3^{j_3} c_4^{j_4} c_5^{j_5}}.
$
The relations for $(j_2, j_3, j_4)<0$ are obtained from symmetry
properties of the integral under consideration: 

$$
(j_1,m_1) \leftrightarrow (j_3,m_3), 
~~~(j_2,m_2) \leftrightarrow (j_4,m_4), 
$$

and 

$$
(j_1,m_1) \leftrightarrow (j_2,m_2), 
~~~(j_3,m_3) \leftrightarrow  (j_4,m_4).
$$

\noindent
$c_5$ from the numerator can be  eliminated by a general
projection-operator method \cite{rec}. Using the decomposition 
$$
k_1 k_2 = A(k_1, k_2, p) + \frac{(k_1 p) (k_2 p)}{p^2}, 
$$

\noindent
where 
$ A(k_1, k_2, p) = k_1^\mu 
\left(\delta_{\mu \nu} - \frac{p_\mu p_\nu}{p^2} \right) k_2^\nu, $
and the property that odd powers of $A(k_1, k_2, p)$ drop out after
integration and for even powers we have 

\begin{eqnarray}
 &\displaystyle \int {\rm d}^N k_1 ~ {\rm d}^N k_2 ~
  f_1[k_1,p]~ f_2[k_2,p]~ A^{2n}(k_1,k_2,p) \,=&
  \nonumber\\*
 &\displaystyle
  \FR { \Gamma(n+\fr 1 2)\, \Gamma \big[ \fr 1 2 (N-1) \big] }
   { \Gamma(\fr 1 2)\, \Gamma \big[ n +\fr 1 2 (N-1) \big] }
  \prod_{j=1}^2 \int {\rm d}^N k_j~ f_j[k_j,p]~ A^n(k_j,k_j,p)
  \, ,& \label{int}
\nonumber 
\end{eqnarray}

\noindent
it is possible to reduce the initial integral to a product of one-loop
integrals.

Using the above relations, F-type integrals with arbitrary
indices are reduced to F-type integrals with only positive
indices or V-type integrals with arbitrary indices. 
For the former case a proper arrangement of recurrence
relations in general reduces the sum of all indices by 1.
These relations are given in appendix A. Only eight diagrams 
({\bf F11111, F00111, F10101, F10110, F01100, F00101, F10100, F00001}) 
with all indices equal to 1 form the basis for F-type diagrams. 

\subsection{V-topology}

Consider now the V-type diagrams. The recurrence relations
we use are $\{ 425\}$, $\{423 \}$ and  the following set: 

\begin{eqnarray}
\{530\} &&  N- 2j_5-j_3
+ \frac{j_5}{c_5} 2 m_5^2
+ \frac{j_3}{c_3} \left( m_3^2 +m_5^2 -m_4^2 + c_4 - c_5 \right)
 = 0,  
\nonumber \\
\{350\} &&  N - 2j_3-j_5 
+ \frac{j_3}{c_3} 2 m_3^2
+ \frac{j_5}{c_5} \left( m_5^2 +m_3^2 -m_4^2 +c_4 - c_3 \right)
 = 0,  
\nonumber \\
\{B\} &&  
\frac{j_5}{c_5}  
\frac{
\left( m_4^2 -m_2^2 +m^2 +c_2-c_4 \right)
\left( m_5^2 -m_3^2 -m_4^2 +c_3+c_4-c_5 \right)}{2 c_6}
\nonumber \\
&&
+ \frac{j_2}{c_2} 2 m^2  
- \left( \frac{j_2}{c_2} + \frac{j_4}{c_4} + \frac{j_5}{c_5}  
\right)
\left( m_4^2 -m_2^2 +m^2 +c_2-c_4 \right)
 = 0,  
\nonumber 
\end{eqnarray}

\noindent 
where $c_6 = c_4 - m_4^2$ (see Ref.\cite{similar}). If $m_4^2 \neq 0$, 
the expression $\frac{1}{c_4 c_6}$ can be simplified later by partial
fraction decomposition.
For all cases $(j_2,j_3,j_5)<0$, except $j_4 < 0$, 
the initial diagram can be reduced to two-loop tadpole-like integrals 
by means of \cite{small1}:

$$
\int d^N k_2 \left( k_2p \right)^{2j} f(k_2,k_1) = 
\frac{(2j)!}{(j)! } 
\left( \frac{p^2}{4} \right)^j
\frac{\Gamma \left(\frac{N}{2} \right)}{ \Gamma \left( \frac{N}{2}+j \right)} 
\int d^N k_2  (k_2^2)^j f(k_2,k_1). $$

\noindent
For $j_4 <0$ we write $c_4 = \overline{c}_4 + m_4^2$ and redefine
$\overline{c}_4  =  c_4 $, which allows to
consider only the massless case.
Then the following recurrence relations are needed:

\begin{enumerate}
\item 
$j_5 \neq 1$

\begin{eqnarray}
\{350\} &&  
\frac{j_5}{c_5} c_4 = 
\frac{j_5}{c_5} \left( c_3 - m_3^2 -m_5^2 \right)
- 2 \frac{j_3}{c_3} m_3^2 - N +2 j_3 +j_5, 
\nonumber 
\end{eqnarray}

\item 
$j_3 \neq 1$
\begin{eqnarray}
\{530\} &&  
\frac{j_3}{c_3} c_4 = 
\frac{j_3}{c_3} \left(c_5 -m_3^2 -m_5^2 \right)
- 2 \frac{j_5}{c_5} m_5^2 - N +2 j_5 +j_3,
\nonumber 
\end{eqnarray}

\item 
$j_2 \neq 1$
\begin{eqnarray}
\{425\} + \{350\} &&  
\frac{j_2}{c_2} c_4 = 
\frac{j_3}{c_3} 2 m_3^2
+ \frac{j_4}{c_4} 2 m_4^2
+ \frac{j_5}{c_5} 2 m_5^2
\nonumber \\
&& 
+ \frac{j_2}{c_2} \left( m_2^2 -m^2 \right)
+ 2N -j_2 -2j_3 -2j_4 -2j_5 .
\nonumber 
\end{eqnarray}

\item
$j_2 = j_3 = j_5 = 1$ 
\begin{eqnarray}
2 B + 2 \{425\} + \{350\}  &&  
 3 N - 4 j_2 -2 j_3 - 2 j_4 - 2 j_5  = 
-2 m_3^2 \frac{j_3}{c_3} - 4 m_2^2 \frac{j_2}{c_2}
\nonumber \\
&& 
+ \frac{j_5}{c_5} \frac{c_2}{c_4} \left( c_4 - c_3 +m_3^2-m_5^2\right)
+ \frac{j_5}{c_5} \frac{c_3}{c_4}  \left(m_2^2-m^2 \right)
\nonumber \\
&& 
+ \left(j_5 + 2j_4 \right) \frac{c_2 + m^2 - m_2^2}{c_4}
+ \frac{j_5}{c_5} \left(m^2 - m_2^2 - 2 m_5^2 \right)
\nonumber \\
&& 
+ \frac{j_5}{c_5} \frac{ \left(m_3^2-m_5^2 \right) \left(m^2-m_2^2 \right)}{c_4}.
\nonumber 
\end{eqnarray}
\end{enumerate}

\noindent
The result of application of the above recurrence relations are V-type
diagrams with only positive indices or J-type integrals with arbitrary
indices. The full set of recurrence relations for the former case is
given in appendix B. The complete set of basic integrals is just given
by {\bf V1111} and {\bf V1001} with indices equal to 1.

\subsection{J-topology}

The integrals of this type are discussed in detail in
Refs.\cite{threshold,Tarasov1}. We only mention here, that to reduce 
the numerator the following recurrence  relation, suggested 
by Tarasov \cite{Tarasov2}, is needed: 

\begin{eqnarray}
(N+\nu_1+\nu_2\ -2) v(\nu_1,\nu_2)=p^2 \{ (\nu_1-1)k_1^2{\bf 1^-}
\!+\nu_2(k_1k_2){\bf 2^-}\} {\bf 1^-} v(\nu_1,\nu_2),
\nonumber
\end{eqnarray}

\noindent 
where 
$ 
v(\nu_1,\nu_2) = 
\int d^Nk_1 d^Nk_2 f(k_1,k_2) (k_1p)^{\nu_1}(k_2p)^{\nu_2}$
and $f(k_1,k_2)$ is an arbitrary scalar function; $\nu_1, \nu_2 >0$
and ${\bf 1^{\pm}}v(\nu_1,\nu_2)\equiv v(\nu_1 \pm 1,\nu_2),etc.$

The master integrals are the following: one prototype {\bf J111} with all indices
equal to 1, and two integrals  of {\bf J011}-type: with indices 111 and 112, 
respectively.

\subsection{Master-integrals}

To obtain the finite part of two-loop physical results one needs
to know the finite part of F-type integrals, V- and J-type
integrals up to order $\varepsilon$, and one-loop integrals up to 
order $\varepsilon^2$. A detailed  discussion of the calculation of 
master-integrals is given in \cite{our}.  Here me mention only,
that  the calculation of the $\varepsilon$ ($\varepsilon^2$) parts
has been performed by the differential equation method
\cite{kotikov}. The results are collected in Appendix C. 

\section{Use of the package}

The package consists of a set of procedures for the calculation of all
two-loop integrals, presented in Fig.\ref{joint} 
(f11111.prc, $\cdots,$ on3.prc, on2.prc, etc),
two-loop tadpoles (vl111.prc, vl011.prc, vl001.prc)
and one-loop integrals (vl1.prc, on1.prc, ons11.prc), 
where ``1''(``0'') in the name of the procedure stands for massive (massless)
lines, respectively. on3 and on2 are two-loop integrals from the SHELL2
package. vl1 is the  one-loop massive bubble. ons11 and on1 denote the
one-loop self-energy on-shell integrals with two and one massive lines, 
respectively. The integration momenta in the package are
denoted by K1 and K2 for two-loop integrals and by K1 for one-loop
integrals, P is the external momentum.  All scalar products in the initial 
diagram must be rewritten in terms of propagators:

\begin{eqnarray}
k_1 p & = & \frac{c_1-c_3+m_3^2-m_1^2-m^2}{2}, 
\nonumber \\
k_2 p & = & \frac{c_2-c_4+m_4^2-m_2^2-m^2}{2}, 
\nonumber \\
k_1 k_2 & = & \frac{c_1+c_2-c_5+m_5^2-m_1^2-m_2^2}{2}, 
\nonumber \\
k_1^2 & = &  c_1-m_1^2, 
\nonumber \\
k_2^2 & = & c_2-m_2^2.
\nonumber 
\end{eqnarray}

\noindent
To specify the type of two- (one-) loop diagrams, the products of
scalar propagators must be substituted by the proper functions of 
F-, V-, J- and ON-type  with arguments  denoting the indices 
and a symbol for the mass shell.

To work with fractions of N-dimensional numbers, two functions, SS and 
NN (originating from the package ``LEO'' \cite{avdeev} ) are used: 

$$
NN(a,j) = \left( N+a \right)^j,~~~~~j > 0, 
$$

\noindent
and

$$
SS(a,j)  = \frac{1}{\left( N+a \right)^j}, ~~~~~j >  0. 
$$

\noindent
After application of each recurrence relation the procedure ``ration'' 
for the simplification of products of SS's and NN's must be called. 
The procedure ``finitem'' substitutes the values of master 
integrals and performs the expansion of the functions NN and SS 
in $\varepsilon$.

The integration procedure starts with F-type integrals. We apply
the recurrence relations given explicitly in appendix A. After  applying
them several times, the integrand is reduced to the master integrals or 
to new, more simple, integrals like V-type, e.g. Then we apply several times 
the recurrence relations of Appendix B for the V-types. One needs to call all
procedures step by step to reduce the initial diagram to the set of master 
integrals. The recommended sequence for calling the procedures is the
following one:

\begin{verbatim}
#call f11111{'TIMES'}
#call f01111{'TIMES'}
#call f11110{'TIMES'}
#call f00111{'TIMES'}
#call f10101{'TIMES'}
#call f10110{'TIMES'}
#call f01100{'TIMES'}
#call f00101{'TIMES'}
#call f10100{'TIMES'}
#call f00100{'TIMES'}
#call f00001{'TIMES'}
#call f00000{'TIMES'}
#call v1111{'TIMES'}
#call v0111{'TIMES'}
#call v1011{'TIMES'}
#call v1010{'TIMES'}
#call v0110{'TIMES'}
#call v1001{'TIMES'}
#call v0010{'TIMES'}
#call v0001{'TIMES'}
#call v0000{'TIMES'}
#call j011{'TIMES'}
#call on3{'TIMES'}
#call on2{'TIMES'}
#call vl111{'TIMES'}
#call vl011{'TIMES'}
#call vl001{'TIMES'}
#call on1{2*'TIMES'}
#call ons11{2*'TIMES'}
#call vl00{'TIMES'}
#call vl1{2*'TIMES'}
\end{verbatim}

All programs of the package are realized with the help of FORM
procedure  facilities. To perform the integration, one needs to call
the procedures with the name of the corresponding prototypes and one
argument which determines how often the recurrence relations are to be 
called. This number depends on the complexity of the calculated
diagram. In most cases it is equal to the sum of indices of the integrand. 

\begin{figure}[ht]
\centerline{\vbox{\epsfysize=50mm \epsfbox{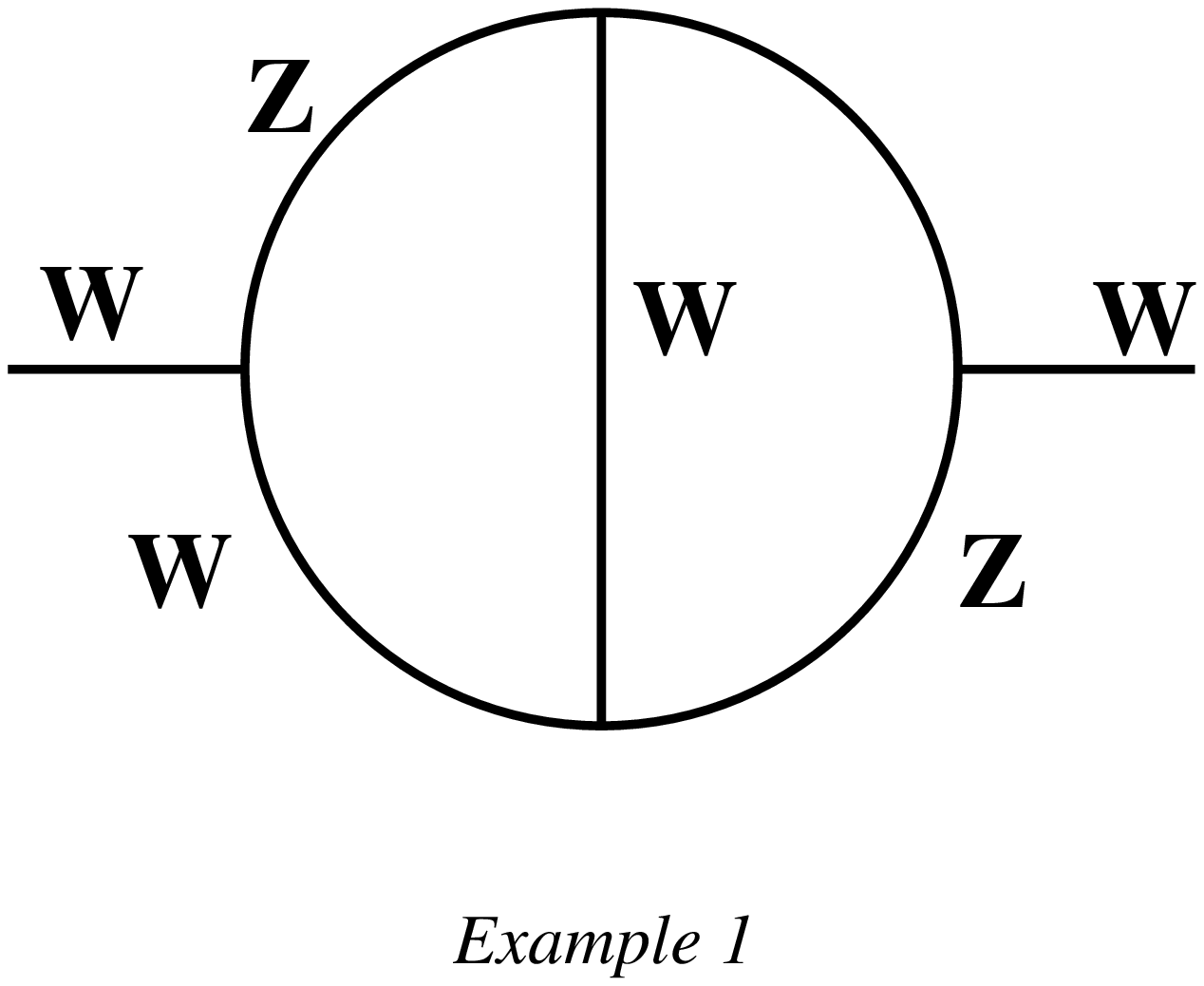}}}
\caption{\label{example} }
\end{figure}

Let us consider as an example, the physical diagram shown in Fig.\ref{example} in more
detail. The FORM input is created automatically by {\it DIANA} \cite{DIANA}. 
For two-loop self-energy diagrams {\it DIANA} generates all necessary information, 
e.g. identifying symbols for the particles of the diagram  and their masses , 
distribution of integration momenta, Feynman rules (linear/nonlinear gauges), 
number of fermion loops, symmetry factors, etc.  To calculate the transverse part 
it is sufficient to make the following substitutions:

\begin{verbatim}
multiply, (d_(mu,nu)-p(mu)*p(nu)/p.p)*SS(-1,1);
.sort
id k1.p = (k1.k1-c3)/2;
id k2.p = (k2.k2-c4+mmZ-mmW)/2;
id k1.k2 = -(c5 - k1.k1 - k2.k2 - mmW)/2;
id k1.k1 = c1 - mmZ;
id k2.k2 = c2 - mmW;
id p.p^j? = (-mmW)^j;
.sort
id mmZ^j? = mmW^j;
id 1/c1^j1?/c2^j2?/c3^j3?/c4^j4?/c5^j5? = F11111(j1,j2,j3,j4,j5,mmW),
\end{verbatim}

\noindent
where we have explicitly set $ m_W^2 =  m_Z^2 $
Calling the above routines (not all of them are needed in this special
case) yields the result

\begin{eqnarray}
&&
{\it Example1}  = 
 m_W^2 {\bf F11111}(1,1,1,1,1,m_W^2) 
\Biggl( 33 - \frac{11}{N-1} \Biggr)
\nonumber \\
&& 
+ {\bf V1111}(1,1,1,1,m_W^2) \Biggl( 
-42 + \frac{2}{N-1} + \frac{27}{4(N-1)^2}
\Biggr)
\nonumber \\
&& 
+ \frac{{\bf J111}(1,1,1,m_W^2)}{m_W^2} \Biggl(
\frac{4}{3} - \frac{26}{3(N-1)} + \frac{17}{4(N-1)^2}
\Biggr)
\nonumber \\
&&
+ \frac{{\bf VL111}(1,1,1,m_W^2)}{m_W^2} \Biggl(
20 - \frac{5}{2(N-1)} - \frac{9}{4(N-1)^2}
\Biggr)
\nonumber \\
&&
+ \left[ {\bf ONS11}(1,1,m_W^2) \right]^2
 \Biggl( \frac{5}{(N-1)} -\frac{21}{2} \Biggr)
\nonumber \\
&&
+ {\bf ONS11}(1,1,m_W^2) 
 \Biggl(
-\frac{17}{3(N-4)} - \frac{1}{(N-2)} + \frac{20}{3(N-1)} -  \frac{6}{(N-1)^2}
\Biggr)
\nonumber \\
&&
+ 
\Biggl( 
\frac{64}{3(N-4)} - \frac{24}{(N-4)^2} - \frac{32}{(N-2)}
- \frac{8}{(N-2)^2} + \frac{32}{3(N-1)} 
\Biggr), 
\nonumber 
\end{eqnarray}

\noindent
where the overall factor is $\frac{g^4 m_W^2}{(16 \pi^2)^2}$.
After calling ``finitem'' we have:

\begin{eqnarray}
{\it Example1}  & = & -\frac{1417}{24 \epsilon^2} 
- \frac{1}{\varepsilon} \Biggl(
\frac{10375}{48} - \frac{665}{12} \frac{\pi}{\sqrt{3}} \Biggr)
- \frac{21187}{32} + \frac{21}{8} \zeta(2) -  \frac{88}{3} \zeta(3)
\nonumber \\
&&
+ \frac{4007}{18}  \frac{\pi}{\sqrt{3}} 
- \frac{665}{12} \frac{\pi}{\sqrt{3}} \ln 3 
+ \frac{16449}{16} S_2 + 132 \frac{\pi}{\sqrt3} S_2 
\nonumber \\
&& 
\approx
- \frac{59.0}{\varepsilon^2} - \frac{115.6}{\varepsilon} -69.6.
\nonumber 
\end{eqnarray}

So far we have considered only the case of merely one non-zero mass.
Of course it is obvious that as further application of our package
we use it for the expansion of diagrams in terms of mass differences.
In general a `standard' expansion of the scalar propagators in terms of 
the mass difference, i.e. in the above case in terms of $ m_W^2 -  m_Z^2 $, 
yields as expansion coefficients again integrals, which can be handled 
by our package. In order to demonstrate this possibility, we extend the
calculation of the diagram in Fig.\ref{example} up to the second order
in $\Delta \equiv 1-m_W^2/m_Z^2   = \sin^2 \theta_W$ with the result

\begin{eqnarray}
&&
{\it Example1}  = 
\frac{1}{\varepsilon^2} \Biggl(
- \frac{1417}{24} + \frac{667}{8}  \Delta - \frac{95}{4} \Delta^2
\Biggr)
\nonumber \\ &&
- \frac{1}{\varepsilon} \Biggl(
\frac{10375}{48} - \frac{24197}{72} \Delta +  \frac{17957}{144} \Delta^2
\Biggr)
+ \frac{1}{\varepsilon} \frac{\pi}{\sqrt{3}} \Biggl(
\frac{665}{12} - \frac{137}{3} \Delta - \frac{635}{72} \Delta^2
\Biggr)
\nonumber \\
&&
+ \frac{\pi}{\sqrt{3}} \Biggl( 
\frac{4007}{18} - \frac{21881}{72} \Delta + \frac{45067}{432} \Delta^2 \Biggr)
- \frac{\pi}{\sqrt{3}} \ln 3 \Biggl( 
\frac{665}{12} - \frac{137}{3} \Delta - \frac{635}{72} \Delta^2
\Biggr)
\nonumber \\
&&
+ S_2 \Biggl( 
\frac{16449}{16} - \frac{7461}{8} \Delta + \frac{2195}{32} \Delta^2
\Biggr)
+ \frac{\pi}{\sqrt{3}} S_2  \Biggl( 
132 - \frac{351}{4} \Delta + \frac{699}{4} \Delta^2
\Biggr)
\nonumber \\
&&
- \Biggl(
\frac{21187}{32} - \frac{216049}{216} \Delta + \frac{403637}{864} \Delta^2
\Biggr)
+ \zeta(2) \Biggl( 
\frac{21}{8} - \frac{1477}{36} \Delta + \frac{23179}{432} \Delta^2
\Biggr)
\nonumber \\
&&
- \zeta(3) \Biggl( 
\frac{88}{3} - \frac{39}{2} \Delta + \frac{233}{6}  \Delta^2
\Biggr) 
\approx
\frac{1}{\varepsilon^2} \left( 
-59.0 + 83.4 \Delta - 23.8  \Delta^2 
\right)
\nonumber \\ && 
+ 
\frac{1}{\varepsilon} 
\left( 
-115.6 + 253.2 \Delta  - 140.7 \Delta^2
\right)
-69.6 + 211.6 \Delta - 118.4 \Delta^2.
\nonumber 
\end{eqnarray}

With the numerical value $\Delta \simeq 0.23$, we see that the convergence
in $\Delta$ of the above series in the three contributions is quite good and
it can be expected that it will even improve due to `gauge cancellations'
if a complete gauge invariant subset of diagrams is taken into account.

\section{Conclusion}

The presented package has been developed for the calculation of two-loop
self energy diagrams with only one non-zero mass. All mass combinations
of the Standard Model with heavy masses are included.
In this sense it is an extension of the existing 
package {\bf SHELL2}, which takes into account only diagrams occurring
in QED and QCD. We have also shown that this package can be
used for the case of different masses in the SM by expanding in terms of
mass differences. Thus we have provided a program for the evaluation
of at least a large class of on-shell two-loop self energy diagrams in the SM. 
The time of calculation of one diagram,
depending of course on the order of expansion in the mass differences, is 
not very large in general.

\vspace{1cm}

{\bf Acknowledgments}
We are grateful to A.~Davydychev, A.~V.~Kotikov, O.~V.~Tarasov,
M.~Tentyukov and O.~Veretin for useful comments. M.K.~'s 
research has been supported by the DFG project FL241/4-1  and in 
part by RFBR $\#$98-02-16923.

\pagebreak

\begin{center}
{\Large\bf Appendix}
\end{center}

\appendix


\section{The set of recurrence relations for F-type integrals}

\noindent
The full set of recurrence relations valid for arbitrary masses and
momenta is given in Ref.\cite{proceding}. Here we present some
rearrangement of these relations which allows to reduce the positive indices 
of the on-shell diagrams with one mass to 0 or 1.

\begin{enumerate}
 \item  {\bf F11111}
  \begin{eqnarray}
4 m^2   \frac{j_1}{c_1} & = & 
  \frac{j_1}{c_1} \left( c_2 - c_3 - c_5 \right)
+ \frac{j_3}{c_3} \left( 3 c_1 + c_4 - c_5 \right)
\nonumber \\
&& + \frac{j_5}{c_5} \left( 3 c_1 - 3 c_2 + c_4 - c_3 \right)
   - N + 4 j_1, 
\nonumber \\
4 m^2   \frac{j_5}{c_5} & = & 
  \frac{j_1}{c_1} \left( 3 c_5 - 3 c_2 - c_3 \right)
+ \frac{j_3}{c_3} \left( 3 c_5 - 3 c_4 - c_1 \right)
\nonumber \\
&& + \frac{j_5}{c_5} \left(c_4 - c_3 + c_2 - c_1 \right)
   - N + 4 j_5. 
\nonumber 
  \end{eqnarray}

\item {\bf F01111}
  \begin{eqnarray}
N - 2 j_1 - j_3 - j_5  & = &
\frac{j_5}{c_5} \left( c_1 - c_2 \right) + \frac{j_3}{c_3} c_1.
\nonumber 
  \end{eqnarray}

\item {\bf F11110}
  \begin{eqnarray}
N - 2 j_5 - j_1 -j_3  & = &
  \frac{j_1}{c_1} \left( c_5 - c_2 \right)
+ \frac{j_3}{c_3} \left( c_5 - c_4 \right).
\nonumber 
   \end{eqnarray}

\item {\bf F00111}
  \begin{eqnarray}
2 m^2   \frac{j_1}{c_1} & = &
\frac{j_1}{c_1} \left( 2 c_5 - 2 c_2 - c_3   \right)
+ \frac{j_3}{c_3} \left( 2 c_5 - 2 c_4 - 3 c_1   \right)
\nonumber \\
&& + \frac{j_5}{c_5} \left( 3 c_2 - 3 c_1 + c_4 - c_3 \right)
   + 2 N - 3 j_3 -5 j_1,
\nonumber \\
2 m^2   \frac{j_3}{c_3} & = &
\frac{j_1}{c_1} c_3  - \frac{j_3}{c_3} c_1 
+ \frac{j_5}{c_5} \left( c_2 - c_1 + c_3 - c_4 \right)
+j_3 -j_1,
\nonumber \\
2 m^2   \frac{j_5}{c_5} & = &
\frac{j_3}{c_3} c_1 + \frac{j_4}{c_4} c_2 
- 2N + 2 j_1 + 2 j_2 + 2 j_5 + j_3 + j_4.
\nonumber 
  \end{eqnarray}

\item {\bf F10101}
 \begin{eqnarray}
m^2 \frac{j_1}{c_1} & = &
\frac{j_3}{c_3} \left( c_5 - c_4   \right)
+ \frac{j_1}{c_1} \left( c_5 - c_2 - c_3   \right)
\nonumber \\
&& + \frac{j_5}{c_5} \left( c_4 - c_3 \right)
   + j_5 -j_3,
\nonumber \\
m^2   \frac{j_2}{c_2} & = &
\frac{j_5}{c_5} \left( c_3  - c_4 \right)
- \frac{j_2}{c_2} c_4
+N -2j_4 -j_2 -j_5,
\nonumber \\
2 m^2   \frac{j_5}{c_5} & = &
\frac{j_2}{c_2} \left( c_5- c_1 \right) 
+ \frac{j_4}{c_4} \left( c_5 - c_3 \right)
- N + 2 j_5 + j_2 + j_4.
\nonumber   
 \end{eqnarray}

\item {\bf F10110}
  \begin{eqnarray}
2 m^2   \frac{j_4}{c_4} & = &
\frac{j_2}{c_2} c_4
- \frac{j_5}{c_5} \left( c_3 - c_4   \right)
   - N + 2 j_4 + j_2 + j_5,
\nonumber \\
m^2 \frac{j_2}{c_2} & = &
\frac{j_2}{c_2} \left( c_1 - c_5 \right)
+ \frac{j_4}{c_4} \left( c_3 - c_5 \right)
+N - 2 j_5 - j_2 - j_4,
\nonumber \\
m^2 \frac{j_5}{c_5} & = &
\frac{j_5}{c_5} \left( c_1 - c_2 \right)
- \frac{j_4}{c_4} c_2 
+ N - 2 j_2 - j_4 - j_5,
\nonumber \\
m^2   \frac{j_1}{c_1} & = &
\frac{j_1}{c_1} \left( c_5 - c_2 \right)
+ \frac{j_3}{c_3} \left( c_5 - c_4 \right)
- N + 2 j_5 + j_1 + j_3,
\nonumber \\
2 m^2   \frac{j_3}{c_3}  & = &
\frac{j_3}{c_3} \left( c_4 - c_5 \right) 
+ \frac{j_5}{c_5} \left( c_3 - c_4 \right) 
+ \frac{j_1}{c_1} \left( c_2 + c_3 - c_5 \right) 
+ j_3 - j_5
\nonumber 
  \end{eqnarray}

\item {\bf F01100}
 \begin{eqnarray}
2 m^2 \frac{j_1}{c_1} & = &
\frac{j_1}{c_1} \left( 2 c_2 - 2 c_5 + c_3   \right)
+ \frac{j_3}{c_3} \left( 2 c_4 - 2 c_5  + c_1   \right)
\nonumber \\
&& + \frac{j_5}{c_5} \left( c_1 - c_2 + c_3 - c_4  \right)
   + j_1 + j_3 -2 j_5,
\nonumber \\
2  m^2   \frac{j_2}{c_2} & = &
\frac{j_2}{c_2} c_4   + \frac{j_4}{c_4} c_2
+ \frac{j_5}{c_5} \left( c_2 - c_1 + c_4 - c_3  \right)
\nonumber \\
&& -2N +2j_5 +3j_2 +3j_4,
\nonumber \\
m^2   \frac{j_5}{c_5} & = &
\frac{j_5}{c_5} \left( c_2 - c_1 \right) 
- \frac{j_3}{c_3} c_1
+ N - 2 j_1 - j_3 - j_5.
\nonumber   
  \end{eqnarray}

\item {\bf F00101}
 \begin{eqnarray}
 m^2 \frac{j_5}{c_5} & = &
\frac{j_5}{c_5} \left( c_1 - c_2  \right)
+\frac{j_3}{c_3} c_1 
   - N  +2 j_1 + j_3 + j_5,
\nonumber \\
m^2   \frac{j_2}{c_2} & = &
\frac{j_5}{c_5} \left( c_3 - c_4 \right)
- \frac{j_2}{c_2} c_4
+ N - 2 j_4 - j_2 - j_5,
\nonumber \\
m^2 \frac{j_4}{c_4} & = &
2 \frac{j_5}{c_5} \left( c_1 - c_2 \right) 
- \frac{j_4}{c_4} c_2 
+ \frac{j_3}{c_3} c_1
+ 2 j_1 - 2j_2 + j_3 - j_4,
\nonumber \\
2 m^2 \frac{j_3}{c_3} & = &
\frac{j_1}{c_1} c_3 
- 2 \frac{j_3}{c_3} c_1  
+ \frac{j_5}{c_5} \left( 2c_2 - 2c_1 + c_3 - c_4 \right) 
+ N - 3 j_1 -j_5, 
\nonumber \\
m^2 \frac{j_1}{c_1} & = &
\frac{j_1}{c_1} \left( c_5 - c_2 - c_3 \right) 
+ \frac{j_3}{c_3} \left(c_5 - c_4 \right) 
+ \frac{j_5}{c_5} \left(c_4 - c_3 \right) 
+ j_5 -j_3. 
\nonumber 
  \end{eqnarray}

\item {\bf F10100}
 \begin{eqnarray}
2 m^2 \frac{j_1}{c_1} & = &
\frac{j_1}{c_1} \left( c_5 - c_2 - c_3   \right)
+ \frac{j_3}{c_3} \left( c_1 - c_4  + c_5   \right)
\nonumber \\
&& + \frac{j_5}{c_5} \left( c_1 - c_2 - c_3 + c_4  \right)
   -N  + 2 j_1 + 2 j_5,
\nonumber \\
2  m^2   \frac{j_2}{c_2} & = &
\frac{j_2}{c_2} \left( c_1 - c_4 - c_5  \right)
+ \frac{j_4}{c_4} \left( c_2 + c_3 - c_5  \right)
+ \frac{j_5}{c_5} \left( c_2 - c_1 + c_3 - c_4 \right)
+ N -2 j_4 -j_5,
\nonumber \\
2 m^2   \frac{j_5}{c_5} & = &
\frac{j_1}{c_1} \left( 3 c_2 + c_3 -3 c_5\right) 
+ \frac{j_3}{c_3} \left( c_1 + 3 c_4 -3 c_5\right) 
\nonumber \\
&& + \frac{j_5}{c_5} \left( c_1 - c_2 +  c_3 - c_4 \right) 
+ N - 4 j_5.
\nonumber 
  \end{eqnarray}

\item {\bf F00100}
 \begin{eqnarray}
N - 2j_1 -j_3 -j_5  & = &
\frac{j_5}{c_5} \left( c_1 - c_2\right) 
+
\frac{j_3}{c_3} c_1.
\nonumber   
  \end{eqnarray}

\item {\bf F00001}
 \begin{eqnarray}
4 m^2 \frac{j_1}{c_1} & = &
\frac{j_1}{c_1} \left( c_5 - c_2 - 3 c_3   \right)
+ \frac{j_3}{c_3} \left( c_1 - c_4  + c_5   \right)
\nonumber \\
&& + \frac{j_5}{c_5} \left( c_1 - c_2 - 3 c_3 + 3 c_4  \right)
   +N  - 4 j_3,
\nonumber \\
4  m^2   \frac{j_5}{c_5} & = &
\frac{j_1}{c_1} \left( c_5 + c_3 - c_2 \right)
+ \frac{j_3}{c_3} \left( c_5 + c_1 - c_4 \right)
\nonumber \\
&& + \frac{j_5}{c_5} \left( c_1 - c_2 + c_3 - c_4 \right)
- 3 N + 4 j_1 + 4 j_3 + 4j_5.
\nonumber 
  \end{eqnarray}

\end{enumerate}


\section{The set of recurrence relations for V-type integrals}


\noindent
The full set of recurrence relations for the V-type integrals
is also given in Ref.\cite{proceding}. Here we give our
rearrangement of these relations which again allows to reduce positive indices 
of the on-shell diagrams with one mass to 0 or 1.

\begin{enumerate}
 \item  {\bf V1111}
  \begin{eqnarray}
3 m^2   \frac{j_3}{c_3} & = & 
  \frac{j_3}{c_3} \left( c_4 - c_5 \right)
+ 2 \frac{j_5}{c_5} \left( c_3 - c_4 \right)
   - N + 3 j_3, 
\nonumber \\
3 m^2   \frac{j_2}{c_2} & = & 
 j_5 \left( \frac{c_2}{c_6} - \frac{c_2 c_3}{c_5 c_6} \right)
+ 2 \frac{j_4}{c_4} c_2 + \frac{j_5}{c_5} c_2 
- \frac{j_2}{c_2} c_4
- N + 3 j_2, 
\nonumber \\
3 m^2   \frac{j_4}{c_4} & = & 
\frac{j_5}{2} \left( \frac{c_2 c_3}{c_5 c_6} - \frac{c_2}{c_6} \right)
+ 2 \frac{j_2}{c_2} c_4 - \frac{j_4}{c_4} c_2 
+ \frac{j_3}{c_3} \left( c_4 -   c_5 \right)
\nonumber \\
&& + \frac{j_5}{2 } \frac{2 c_4 - 2 c_3 - c_2}{c_5}
- \frac{N}{2} + 3 j_4.
\nonumber 
  \end{eqnarray}

\item {\bf V0111}
  \begin{eqnarray}
3 m^2   \frac{j_3}{c_3} & = &
  \frac{j_3}{c_3} \left( c_4 - c_5 \right)
+ 2 \frac{j_5}{c_5} \left( c_3 - c_4 \right)
   - N + 3 j_3,
\nonumber \\
2 m^2   \frac{j_4}{c_4} & = &
\frac{2}{3} \frac{j_3}{c_3} \left(c_4 - c_5 \right)
+ \frac{2}{3} \frac{j_5}{c_5} \left(c_4 - c_3 \right)
+ \frac{j_2}{c_2} c_4 
\nonumber \\
&& - \frac{2}{3} N + 2 j_4 + j_2, 
\nonumber \\
2 \left(N- 2 j_2 - j_4 \right) & = &
\frac{j_5}{c_5} \left( c_2 + c_4 \right)
+ 2 \frac{j_4 }{c_4} c_2
+ j_5 \frac{c_5 - c_3 }{c_5} \frac{ c_2 + c_4}{c_6}.
\nonumber 
  \end{eqnarray}

\item {\bf V1011}
  \begin{eqnarray}
2 m^2   \frac{j_3}{c_3} & = &
\frac{j_5}{c_5} \left( c_3 - c_4 \right)  - N + 2 j_3 + j_5,
\nonumber \\
\left(N -  2 j_5 - j_3 \right) & = &
 \frac{j_3 }{c_3} \left( c_5 - c_4 \right).
\nonumber 
   \end{eqnarray}

\item {\bf V1010}
  \begin{eqnarray}
2 m^2   \frac{j_3}{c_3} & = &
\frac{j_2}{c_2} c_4 
- 2 N + j_2 + 2j_3 + 2 j_4 + 2j_5,
\nonumber \\
m^2  \frac{j_5}{c_5} & = &
\frac{j_5}{c_5} \left( c_3 -c_4 \right) - \frac{j_2}{c_2} c_4 
+ N -2 j_4 - j_2 - j_5,
\nonumber \\
4 m^2   \frac{j_2}{c_2} & = &
m^2 \frac{j_5}{c_5} \frac{c_2}{c_4} - \frac{j_2}{c_2} c_4 
+ \left( j_5 + 2 j_4 \right) \frac{c_2}{c_4}
\nonumber \\
&& 
- \frac{j_5}{c_5} \frac{c_2}{c_4} \left( c_3 - c_4 \right) 
- N + 3 j_2, 
\nonumber \\
4 \frac{m^2}{c_4} \left( j_2 + 2 j_4 \right) & = &
- m^2 \frac{j_5}{c_5} \frac{c_2}{c_4}
+  \frac{j_2}{c_2}  \left(4 c_3 - 3 c_4 - 4 c_5 \right)
- \left( j_5 + 2 j_4 \right) \frac{c_2}{c_4}
\nonumber  \\
&&
+ 4 \left( j_2 + 2 j_4 \right) \frac{c_3 - c_5}{c_4}
+ \frac{j_5}{c_5} \frac{c_2}{c_4} \left( c_3 - c_4 \right)
\nonumber  \\
&&
+ 9 N - 16 j_5 - 8 j_4 -7 j_2,
\nonumber \\
2 j_4 + j_2 - 2 j_5 & = & 
2 \frac{j_3}{c_3} \left( c_5 - c_4 \right) - \frac{j_2}{c_2} c_4.
\nonumber 
  \end{eqnarray}

\item {\bf V0110}
 \begin{eqnarray}
m^2 \frac{j_2}{c_2} & = &
\frac{j_5}{c_5} \left( c_3 - c_4   \right)
- \frac{j_2}{c_2} c_4 
+ N - 2j_4  -j_2 - j_5, 
\nonumber \\
4 m^2  \frac{j_5}{c_5} & = &
m^2 \frac{j_2}{c_2} \frac{c_5 - c_3}{c_4}
+ \frac{j_2}{c_2} \left( c_5 - c_3 \right)
+ \frac{j_5}{c_5} \left( c_3 - c_4 \right)
\nonumber \\
&& + \left( j_2 + 2 j_4 \right) \frac{c_5 - c_3 }{c_4}
- N + 3 j_5, 
\nonumber \\
4 \frac{m^2}{c_4} \left( j_5 + 2 j_4 \right) & = &
4 m^2  \frac{j_5}{c_5} \frac{c_3}{c_4}
+ m^2 \frac{j_2}{c_2} \frac{ c_3 - c_5}{c_4}
+ \frac{j_2}{c_2} \left( c_3 - c_5 \right)
- 4 j_5 \frac{c_2}{c_4 } 
\nonumber \\
&& +  j_2 \frac{ c_3 - c_5}{c_4}
- 2 \frac{j_4}{c_4} \left( 4 c_2 - c_3 + c_5 \right)
\nonumber \\
&& + \frac{j_5}{c_5} \left( 4 \frac{c_2 c_3}{c_4 } 
- 4 c_2 + 3 c_3 - 3 c_4  \right)
\nonumber \\
&&
+ 9 N - 16 j_2 - 8 j_4 - 7 j_5.
\nonumber   
 \end{eqnarray}

\item {\bf V1001}
 \begin{eqnarray}
\frac{j_3}{c_3}  & = &
- \frac{ \left( N - 2j_3 - 2 j_5  \right)  \left( N - j_3 - j_5 - 1 \right)}
{\left( N  -2 j_3 -2 \right) c_6},
\nonumber \\
3 m^2 \frac{j_4}{c_4} & = &
2 \frac{j_2}{c_2} c_4 - \left( \frac{j_4}{c_4} + \frac{j_6}{c_6} \right) c_2
- N  + 3 j_4 + 3 j_6,
\nonumber  \\
3 m^2 \frac{j_2}{c_2} & = &
2 \left( \frac{j_4}{c_4} + \frac{j_6}{c_6} \right) c_2
- \frac{j_2}{c_2} c_4  - N  + 3 j_2.
\nonumber  
  \end{eqnarray}

\item {\bf V0010}
 \begin{eqnarray}
m^2 \frac{j_3}{c_3} & = &
\frac{j_3}{c_3} \left( c_5 - c_4 \right)
- N  + 2 j_5 + j_3, 
\nonumber \\
m^2   \frac{j_5}{c_5} & = &
2 \frac{j_3}{c_3} \left( c_4 - c_5 \right)  
+ \frac{j_5}{c_5} \left( c_3 - c_4 \right)
+ N - 3 j_5,
\nonumber \\
m^2   \frac{j_2}{c_2} & = &
2 \frac{j_3}{c_3} \left( c_5 - c_4 \right) 
- \frac{j_2}{c_2} c_4
+ 2 j_5 - 2j_4 - j_2,
\nonumber  \\
\frac{m^2}{c_4} \left( N + 2 j_4 - 2 j_5 \right) & = &
2 m^2 \frac{j_3}{c_3} \frac{c_5}{c_4} 
- \left( 2 j_4 + j_5 \right) \frac{ c_2}{c_4}
+ \frac{j_5}{c_5} \frac{c_2}{c_4}   \left( c_3 - c_4 \right)
\nonumber  \\
&& 
- m^2 \frac{j_5}{c_5} \frac{c_2}{c_4}
+ 3 N - 4 j_2 - 2 j_3 - 2 j_4 - 2 j_5.
\nonumber    
  \end{eqnarray}

\item {\bf V0001}
 \begin{eqnarray}
\frac{j_3}{c_3} & = &
- \frac{ \left( N - 2j_3 - 2 j_5  \right)  \left( N - j_3 - j_5 - 1 \right)}
{\left( N  -2 j_3 -2 \right) c_6 },
\nonumber \\
N - 2 j_2 - j_4  & = &
\frac{j_4}{c_4} c_2 
- \left( \frac{N}{2} -2 - j_6\right) \frac{c_2 + c_4 }{c_6}.
\nonumber 
  \end{eqnarray}

\end{enumerate}


\section{Analytical results}

\noindent
For completeness, we present here the analytical results for all two-loop 
integrals shown in Fig.\ref{joint} 

\begin{eqnarray}
{\bf F11111}(1,1,1,1,1,m) &  = & - \zeta(3) + \frac{9}{2} \frac{\pi}{\sqrt{3}} S_2
+ {\cal O} (\varepsilon),
\nonumber \\
{\bf F01111}(1,1,1,1,1,m) &  = & 
\frac{4}{3} \zeta(2) \ln 3 - \frac{5}{3}\zeta(3) 
+ 3 \frac{\pi}{\sqrt{3}} S_2 + {\cal O} (\varepsilon),
\nonumber \\
{\bf F11110}(1,1,1,1,1,m)  &  = & 
- \frac{4}{3} \zeta(2) \ln 3 + \frac{2}{3}\zeta(3)
+ 6  \frac{\pi}{\sqrt{3}} S_2 + {\cal O}
(\varepsilon),
\nonumber \\
{\bf F00111}(1,1,1,1,1,m)  &  = & 9  \frac{\pi}{\sqrt{3}} S_2 + {\cal O}
(\varepsilon),
\nonumber \\
{\bf F10101}(1,1,1,1,1,m)  &  = & -4 \zeta(3) + \frac{27}{2}
\frac{\pi}{\sqrt{3}} S_2 + i \frac{\pi}{3} \zeta(2)
+ {\cal O} (\varepsilon),
\nonumber \\
{\bf F10110}(1,1,1,1,1,m)
&  = & - \zeta(3) + 9 \frac{\pi}{\sqrt{3}} S_2 + {\cal O} (\varepsilon),
\nonumber \\
{\bf F01101}(1,1,1,1,1,m)  &  = & 
6 \zeta (2) \ln 2 - \frac{3}{2} \zeta (3)
+ {\cal O} (\varepsilon),
\nonumber \\
{\bf F01100}(1,1,1,1,1,m)
&  = & \frac{27}{2} \frac{\pi}{\sqrt{3}} S_2  + i \pi \zeta(2)
+ {\cal O} (\varepsilon),
\nonumber \\
{\bf F00101}(1,1,1,1,1,m) &  = &  - 3 \zeta(3) + \frac{27}{2}
\frac{\pi}{\sqrt{3}} S_2
+ i \pi \zeta(2) + {\cal O} (\varepsilon),
\nonumber \\
{\bf F10100}(1,1,1,1,1,m) &  = &
- 2 \zeta(3) + 9 \frac{\pi}{\sqrt{3}} S_2 + i \frac{2}{3} \pi \zeta(2)
+ {\cal O} (\varepsilon),
\nonumber \\
{\bf F00110}(1,1,1,1,1,m)  &  = & 
- \frac{\zeta(2)}{\varepsilon} - 2 \zeta(2)
- 2 \zeta(3)  + {\cal O} (\varepsilon),
\nonumber \\
{\bf F00100}(1,1,1,1,1,m)
&  = & 
- 3 \zeta(3) - 2 i \pi \zeta(2) + {\cal O} (\varepsilon),
\nonumber \\
{\bf F00001}(1,1,1,1,1,m) &  = &
- 3 \zeta(3) + i \pi \zeta(2) + {\cal O} (\varepsilon),
\nonumber \\
{\bf F00000}(1,1,1,1,1,m)
&  = & - 6 \zeta(3) + {\cal O} (\varepsilon),
\nonumber 
\end{eqnarray}

\noindent
where 

\begin{eqnarray}
{\bf F}\{ {\cal ABIJK} \}  (a,b,i,j,k,m)
& \equiv &  m^2 K^{-2}
\int  d^Nk_1 d^Nk_2
\nonumber \\
&&
P^{(a)}(k_1,{\cal A} m)
P^{(b)}(k_2,{\cal B} m)
\nonumber \\
&&
P^{(i)}(k_1-p,{\cal I}m)
P^{(j)}(k_2-p,{\cal J}m)
\nonumber \\
&&
\left. P^{(k)}(k_1-k_2,{\cal K}m) \right|_{p^2=-m^2},
\nonumber
\end{eqnarray}

\noindent 
and  
$$
K = \frac{\Gamma(1+\varepsilon)}{ \left(4 \pi \right)^{\frac{N}{2}}
\left( m^2 \right)^{\varepsilon}}, ~~~
P^{(l)}(k,m) \equiv \frac{1}{(k^2+m^2)^l},$$
and where the normalization factor $1/(2 \pi)^N$ for each loop is assumed;
${\cal A,B,I,J,K} = 0,1$ and 
$$S_2 = \frac{4}{9\sqrt{3}} Cl_2 \left(\frac{\pi}{3} \right)=0.2604341376
\cdots.$$

\begin{eqnarray}
&& {\bf V1111}(1,1,1,1,m) = \frac{1}{2 \varepsilon^2}  
+ \frac{1}{\varepsilon} 
\left( \frac{5}{2} - \frac{\pi}{\sqrt{3}} \right)
+ \frac{19}{2} - \frac{\zeta(2)}{2} - 4 \frac{\pi}{\sqrt{3}}
- \frac{63}{4} S_2 
\nonumber \\
&& 
+ \frac{\pi}{\sqrt{3}} \ln 3
+ \varepsilon \Biggl\{
\frac{65}{2} - 6 \zeta(2) - \frac{9}{2} \zeta(3) 
-12 \frac{\pi}{\sqrt{3}}
- 63 S_2 
+ 4 \zeta(2) \ln3 
+ 9 S_2 \frac{\pi}{\sqrt{3}} 
\nonumber \\
&&
+ \frac{63}{4} S_2 \ln3 
+ 4  \frac{\pi}{\sqrt{3}} \ln 3
- \frac{1}{2}\frac{\pi}{\sqrt{3}} \ln^2 3
- \frac{9}{2} \frac{\pi}{\sqrt{3}} \zeta (2) 
- \frac{21}{2} \frac{Ls_3 \left(\frac{2\pi}{3} \right)}{\sqrt{3}} 
\Biggr\}
+ {\cal O} (\varepsilon^2),
\nonumber \\
&& {\bf V0111}(1,1,1,1,m) = 
\frac{1}{2 \varepsilon^2} + \frac{5}{2 \varepsilon}
+ \frac{19}{2} - \frac{2}{3} \zeta(2) - \frac{\pi}{\sqrt{3}}
- \frac{27}{2} S_2
\nonumber \\
&&
+ \varepsilon \Biggl\{
\frac{65}{2} - \frac{8}{3} \zeta(2) 
+ \frac{2}{3} \zeta(3)
-7 \frac{\pi}{\sqrt{3}}
- \frac{135}{2} S_2
+ \frac{27}{2} S_2 \ln3
- 3 \frac{\pi}{\sqrt{3}} \zeta(2)
\nonumber \\
&&
+ 3 \frac{\pi}{\sqrt{3}} \ln 3 
- 9 \frac{Ls_3 \left(\frac{2\pi}{3} \right)}{\sqrt{3}}
\Biggr\}
+ {\cal O} (\varepsilon^2),
\nonumber \\
&& {\bf V1011}(1,1,1,1,m) =
\frac{1}{2 \varepsilon^2} + \frac{1}{\varepsilon}
\left(\frac{5}{2} - \frac{\pi}{\sqrt{3}} \right)
+ \frac{19}{2} + \frac{\zeta(2)}{3} - 5 \frac{\pi}{\sqrt{3}}
- 9 S_2 
\nonumber \\
&&
+ \frac{\pi}{\sqrt{3}} \ln 3
+ \varepsilon \Biggl\{
\frac{65}{2} + \frac{4}{3} \zeta(2)
- \frac{\zeta(3)}{3}
- 19 \frac{\pi}{\sqrt{3}}
- \frac{99}{2} S_2
+ 9 S_2 \ln3
- 3 \frac{\pi}{\sqrt{3}} \zeta(2)
\nonumber \\
&&
+ 7 \frac{\pi}{\sqrt{3}} \ln 3
- \frac{1}{2} \frac{\pi}{\sqrt{3}} \ln^2 3
- 6 \frac{Ls_3 \left(\frac{2\pi}{3} \right)}{\sqrt{3}}
\Biggr\}
+ {\cal O} (\varepsilon^2),
\nonumber \\
&& {\bf V1110}(1,1,1,1,m) =
\frac{1}{2 \varepsilon^2} + \frac{5}{2 \varepsilon}
+ \frac{19}{2} - 4 \zeta(2)
\nonumber \\
&&
+ \varepsilon \Biggl\{
\frac{65}{2} - 20 \zeta(2) - 14 \zeta(3)
+ 24 \zeta(2) \ln 2
\Biggr\} + {\cal O} (\varepsilon^2),
\nonumber \\
&& {\bf V1010}(1,1,1,1,m) =
\frac{1}{2\varepsilon^2} + \frac{5}{2 \varepsilon}
+ \frac{19}{2} + \zeta(2) - 3 \frac{\pi}{\sqrt{3}}
\nonumber \\
&&
+ \varepsilon \Biggl\{
\frac{65}{2} + 2 \zeta(2) - \zeta(3)
- 15 \frac{\pi}{\sqrt{3}}
- \frac{81}{2} S_2
+ 9 \frac{\pi}{\sqrt{3}} \ln 3
\Biggr\}
+ {\cal O} (\varepsilon^2),
\nonumber \\
&& {\bf V0110}(1,1,1,1,m) =
\frac{1}{2 \varepsilon^2} + \frac{1}{\varepsilon} \left(\frac{5}{2} - i \pi \right)
+ \frac{19}{2} - 5 \zeta(2) - 3 \frac{\pi}{\sqrt{3}} - 2 i \pi 
\nonumber \\
&&
+ \varepsilon \Biggl\{
\frac{65}{2} - 10 \zeta(2) - \zeta(3)
- 15 \frac{\pi}{\sqrt{3}}
- \frac{81}{2} S_2
+ 9 \frac{\pi}{\sqrt{3}} \ln 3
+ 2 i \pi \left[ \zeta (2) -2 \right]
\Biggr\}
+ {\cal O} (\varepsilon^2),
\nonumber \\
&& 
{\bf V1001}(1,1,1,1,m) = \frac{1}{2 \varepsilon^2}  
+ \frac{1}{\varepsilon} 
\left( \frac{5}{2} - \frac{\pi}{\sqrt{3}} \right)
+ \frac{19}{2} + \frac{3}{2} \zeta(2) - 4 \frac{\pi}{\sqrt{3}}
- \frac{63}{4} S_2 
\nonumber \\
&&
+ \frac{\pi}{\sqrt{3}} \ln 3
+ \varepsilon \Biggl\{
\frac{65}{2} + 8 \zeta(2) + \frac{3}{2} \zeta(3)
-12 \frac{\pi}{\sqrt{3}}
- 63 S_2
+ \frac{63}{4} S_2 \ln3
\nonumber \\
&&
+ 4  \frac{\pi}{\sqrt{3}} \ln 3
- \frac{1}{2}\frac{\pi}{\sqrt{3}} \ln^2 3
- \frac{21}{2} \frac{\pi}{\sqrt{3}} \zeta (2)
- \frac{21}{2} \frac{Ls_3 \left(\frac{2\pi}{3} \right)}{\sqrt{3}}
\Biggr\}
+ {\cal O} (\varepsilon^2),
\nonumber \\
&& {\bf V0011}(1,1,1,1,m) =
\frac{1}{2 \varepsilon^2} + \frac{5}{2 \varepsilon} 
+ \frac{19}{2} - 2 \zeta(2) 
+ \varepsilon \Biggl\{ \frac{65}{2} - 6\zeta(2) - 4 \zeta(3) \Biggr\}
+ {\cal O} (\varepsilon^2),
\nonumber \\
&& {\bf V0010}(1,1,1,1,m) =
\frac{1}{2 \varepsilon^2} + \frac{1}{\varepsilon} \left(\frac{5}{2} - i \pi \right)
+ \frac{19}{2} - 7 \zeta(2) -  3 i \pi 
\nonumber \\
&&
+ \varepsilon \Biggl\{
\frac{65}{2} - 17 \zeta(2) - 11 \zeta(3)
+ i \pi \left[ 2 \zeta (2) -7 \right] 
\Biggr\}
+ {\cal O} (\varepsilon^2),
\nonumber \\
&& {\bf V1000}(1,1,1,1,m) =
\frac{1}{2 \varepsilon^2} + \frac{5}{2 \varepsilon}
+ \frac{19}{2} +  2 \zeta(2) 
+ \varepsilon \Biggl\{
\frac{65}{2} + 10 \zeta(2) + 4 \zeta(3)
\Biggr\} + {\cal O} (\varepsilon^2), 
\nonumber \\
&& {\bf V0001}(1,1,1,1,m) =
\frac{1}{2 \varepsilon^2} + \frac{5}{2 \varepsilon} 
+ \frac{19}{2} +  \zeta(2) -  i \pi 
+ \varepsilon \Biggl\{
\frac{65}{2} - 3 \zeta(2) -  \zeta(3)
 - 7 i \pi 
\Biggr\} 
\nonumber \\
&& 
+ {\cal O} (\varepsilon^2),
\nonumber \\
&& {\bf V0000}(1,1,1,1,m) =
\frac{1}{2 \varepsilon^2} + \frac{1}{\varepsilon} \left(\frac{5}{2} - i \pi \right)
+ \frac{19}{2} - 7 \zeta(2) - 5 i \pi 
\nonumber \\
&&
+ \varepsilon \Biggl\{
\frac{65}{2} - 35 \zeta(2) - 5 \zeta(3)
+ i \pi \left[ 6 \zeta (2) - 19 \right]
\Biggr\}
+ {\cal O} (\varepsilon^2),
\nonumber 
\end{eqnarray}

\noindent

\begin{eqnarray}
{\bf V} \{ {\cal IJKL} \}  (i,j,a,b,m)  & \equiv & K^{-2}
\int  d^Nk_1 d^Nk_2 P^{(i)}(k_2-p,{\cal I}m)
\nonumber \\
&&
P^{(j)}(k_1-k_2,{\cal J}m)
\nonumber \\
&&
P^{(a)}(k_1,{\cal A}m)
\left. 
P^{(b)}(k_2,{\cal B}m)
\right|_{p^2=-m^2},  
\nonumber
\end{eqnarray}

\noindent
and 
$${\rm Ls}_3 \left(\frac{2\pi}{3} \right) = -2.14476721256949 \cdots,$$
where we use the following definition \cite{Lewin}:
$$Ls_3(x) = -\int_0^x \ln^2 \left| 2 \sin \frac{\theta}{2} \right| 
d\theta. $$
To our knowledge ${\rm Ls}_3 \left(\frac{2\pi}{3}\right)$
has appeared for the first time in the calculation of the
$\varepsilon$-part of two-loop tadpole integrals in  Ref.\cite{epsilon}.

\begin{eqnarray}
&& {\bf J111}(1,1,1,m) = - m^2 \Biggl(
\frac{3}{2\varepsilon^2} + \frac{17}{4 \varepsilon} + \frac{59}{8}
+ \varepsilon \Biggl\{\frac{65}{16} + 8 \zeta(2) \Biggr \}
\nonumber \\
&&
- \varepsilon^2 \Biggl\{
\frac{1117}{32} - 52 \zeta(2) + 48 \zeta (2) \ln 2 - 28 \zeta (3)
\Biggr\} + {\cal O} (\varepsilon^3)
\Biggr),
\nonumber \\
&& {\bf J011}(1,1,2,m) = \frac{1}{2 \varepsilon^2} + \frac{1}{2 \varepsilon}
- \frac{1}{2} - \frac{\zeta(2)}{3} + \frac{\pi}{\sqrt{3}}
+ \varepsilon \Biggl\{
- \frac{11}{2} - \frac{2}{3} \zeta(2) 
\nonumber \\
&& 
+ 5 \frac{\pi}{\sqrt{3}} + \frac{\zeta(3)}{3} 
-3  \frac{\pi}{\sqrt{3}} \ln 3
+ \frac{27}{2} S_2 \Biggr\}  
+ \varepsilon^2 \Biggl\{
- \frac{49}{2} - \frac{4}{3} \zeta(2)
+ 19 \frac{\pi}{\sqrt{3}} 
\nonumber \\
&& 
+ \frac{2}{3} \zeta(3)
- 15 \frac{\pi}{\sqrt{3}} \ln 3
+ \frac{135}{2} S_2 
- \frac{81}{2}S_2 \ln 3
+ \frac{9}{2} \frac{\pi}{\sqrt{3}} \ln^2 3 
\nonumber \\
&& 
+ 14 \frac{\pi}{\sqrt{3}} \zeta(2)
+ 27 \frac{Ls_3 \left(\frac{2\pi}{3} \right)}{\sqrt{3}} 
- \frac{3}{2} \zeta(4) \Biggr\}
+ {\cal O} (\varepsilon^3),
\nonumber \\
&& {\bf J011}(1,1,1,m) = -  m^2  
\Biggl( \frac{1}{\varepsilon^2} + \frac{11}{4\varepsilon}
+ \frac{35}{8} + \frac{3}{2} \frac{\pi}{\sqrt{3}}
+ \varepsilon \Biggl\{
\frac{5}{16} + \frac{39}{4} \frac{\pi}{\sqrt{3}} 
\nonumber \\
&& 
-  \frac{9}{2}  \frac{\pi}{\sqrt{3}} \ln 3
+ \frac{81}{4} S_2 \Biggr\}  
- \varepsilon^2 \Biggl\{
\frac{1033}{32} - \frac{1053}{8} S_2 -\frac{345}{8} \frac{\pi}{\sqrt{3}} 
+ \frac{243}{4}S_2 \ln 3
\nonumber \\
&& 
+ \frac{117}{4} \frac{\pi}{\sqrt{3}} \ln 3 
-\frac{27}{4} \frac{\pi}{\sqrt{3}} \ln^2 3
- 21 \frac{\pi}{\sqrt{3}} \zeta(2)
- \frac{81}{2}  \frac{Ls_3 \left(\frac{2\pi}{3} \right)}{\sqrt{3}} 
\Biggr\}
+ {\cal O} (\varepsilon^3)
\Biggr),
\nonumber  \\
&& {\bf J001}(1,1,1,1,m) = -m^2 \Biggl( 
\frac{1}{2 \varepsilon^2} + \frac{5}{4 \varepsilon} 
+ \frac{11}{8} + 2 \zeta(2) 
- \varepsilon \Biggl\{
\frac{55}{16} - 5 \zeta(2) - 4 \zeta(3)
\Biggr\}
\nonumber \\
&&
- \varepsilon^2 \Biggl\{
\frac{949}{32} - \frac{11}{2} \zeta(2) - 10 \zeta(3)
- 32 \zeta(4)
\Biggr\}
+ {\cal O} (\varepsilon^3) \Biggr),
\nonumber \\
&& {\bf J000}(1,1,1,1,m) = m^2 \Biggl( 
\frac{1}{4 \varepsilon} + \frac{13}{8} - \frac{i \pi}{2}
+ \varepsilon \Biggl\{
\frac{115}{16} - \frac{7}{2} \zeta(2) -  \frac{13}{4} i \pi
\Biggr\}
\nonumber \\
&& 
+ \varepsilon^2 \Biggl\{
\frac{865}{32} - \frac{91}{4} \zeta(2) - \frac{5}{2} \zeta(3)
+ i \pi \left[ 3 \zeta(2) - \frac{115}{8} \right]
\Biggr\}
+ {\cal O} (\varepsilon^3)
\Biggr),
\nonumber 
\end{eqnarray}

\noindent
where 

\begin{eqnarray}
{\bf J} \{ {\cal IJL} \}  (i,j,l,m)  & \equiv & K^{-2}
\int  d^Nk_1 d^Nk_2 P^{(i)}(k_2,{\cal I}m)
\nonumber \\
&&
P^{(j)}(k_1-k_2,{\cal J}m) P^{(l)}(k_1-p,{\cal L}m)
\left. \right|_{p^2=-m^2},
\nonumber
\end{eqnarray}

\noindent
and for the $\varepsilon^2$-part of the ${\bf J011}(1,1,2,m)$ integral we take
into account the  ``guessed''  expansion from Ref.\cite{proceding}. 
The expansion  of J111(1,1,1,m) up to $\varepsilon^4$ terms is given
in Ref.\cite{j111}. A very elegant $\varepsilon$-expansion up to arbitrary
order for the one-loop propagator type integral can be found in Ref.\cite{one}.

\end{document}